\documentclass{PoS}

\usepackage{wrapfig}

\title{Cosmic Ray Signatures from Decaying Gravitino Dark Matter}

\ShortTitle{Cosmic Ray Signatures from Decaying Gravitino Dark Matter}

\author{\speaker{N.-E.~Bomark}\\%
        University of Bergen, Norway\\
        E-mail: \email{nils-erik.bomark@ift.uib.no}}

\author{S.~Lola\\
        University of Patras, Greece\\}
\author{P.~Osland\\
        University of Bergen, Norway\\}
\author{A.R.~Raklev\\
        University of Cambridge, UK\\}

      \abstract{We study the charged cosmic rays arising from the slow
        decay of gravitino dark matter within supersymmetric scenarios
        with trilinear R-parity violation. It is shown that operators
        of the $LL\bar E$ type can very well account for the recent
        anomalies in cosmic ray electron and positron data reported by
        PAMELA, ATIC and Fermi LAT, without violating any other
        bounds. This scenario will soon be tested by the Fermi LAT
        data on diffuse gamma ray emission.}

\FullConference{European Physical Society Europhysics Conference on High Energy Physics\\
                 July 16-22, 2009\\
                 Krakow, Poland}

\begin{document}

\section{Introduction}
Among all dark matter candidates, the most studied ones are clearly
the WIMPs, which are particles with electroweak scale masses and
annihilation cross sections.  In supersymmetric models, the lightest
supersymmetric particle (LSP) is a prominent dark matter candidate
when it is protected from decay via R-parity, a symmetry first
introduced to guarantee the stability of the proton.  However, other
symmetries could also provide such protection.

Attempts at detecting dark matter have focused on WIMP properties,
both in direct detection experiments and when using indirect methods,
i.e. the search for annihilation products.  At the same time there is
no direct empirical evidence in favour of the WIMPs; in fact, we know
little about the interaction strength of dark matter.\bigskip

Gravitinos as dark matter have very different features compared to
conventional WIMPs, such as the neutralino, since they do not
annihilate at any measurable rate and are not expected to show up in
direct detection experiments at all. In fact, gravitino dark matter
does not require R-parity; it has been shown that even if this
symmetry is violated, the lifetime of the gravitino can easily be
larger than the age of the universe \cite{TY}.

We will here report on a study \cite{BLOR-pamela} of supersymmetric
models with a gravitino LSP and trilinear R-parity violating operators
(for bilinear R-parity violation see \cite{Chpart}):
\begin{equation}
\lambda L_{i}L_{j}{\bar{E}}_{k}
+\lambda ^{\prime }L_{i}Q_{j}{\bar{D}_{k}}
+\lambda ^{\prime \prime }{\bar{U}_{i}}{\bar{D}_{j}}{\bar{D}_{k}},
\label{Rviol}
\end{equation}
where $L(Q)$ are the left-handed lepton (quark) doublet superfields,
and ${\bar{E}}$ (${\bar{D}},{\bar{U}}$) are the corresponding
left-handed singlet fields. Earlier work \cite{BLOR} studied the gamma
radiation from gravitino decays induced by the operators of
Eq.~(\ref{Rviol}). This is now extended to include charged cosmic
rays.

\section{Charged Particles}

\begin{figure}[!ht]
\begin{centering}
  \includegraphics[width=14cm]{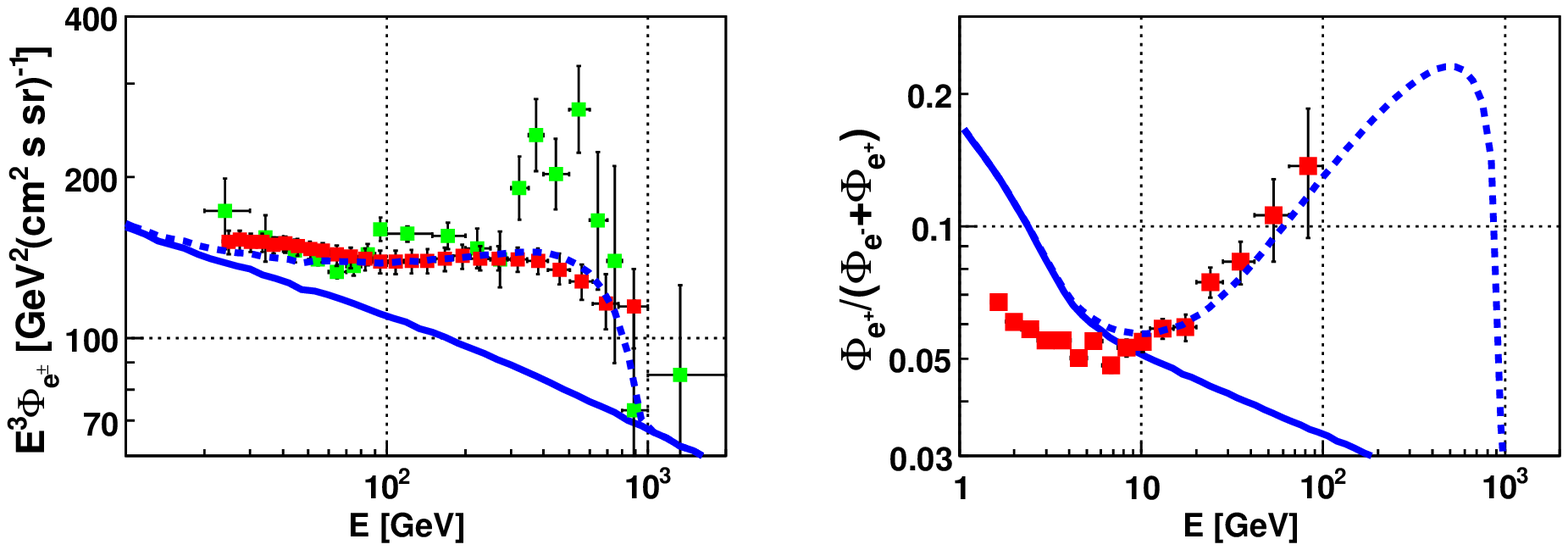}\\
  \includegraphics[width=14cm]{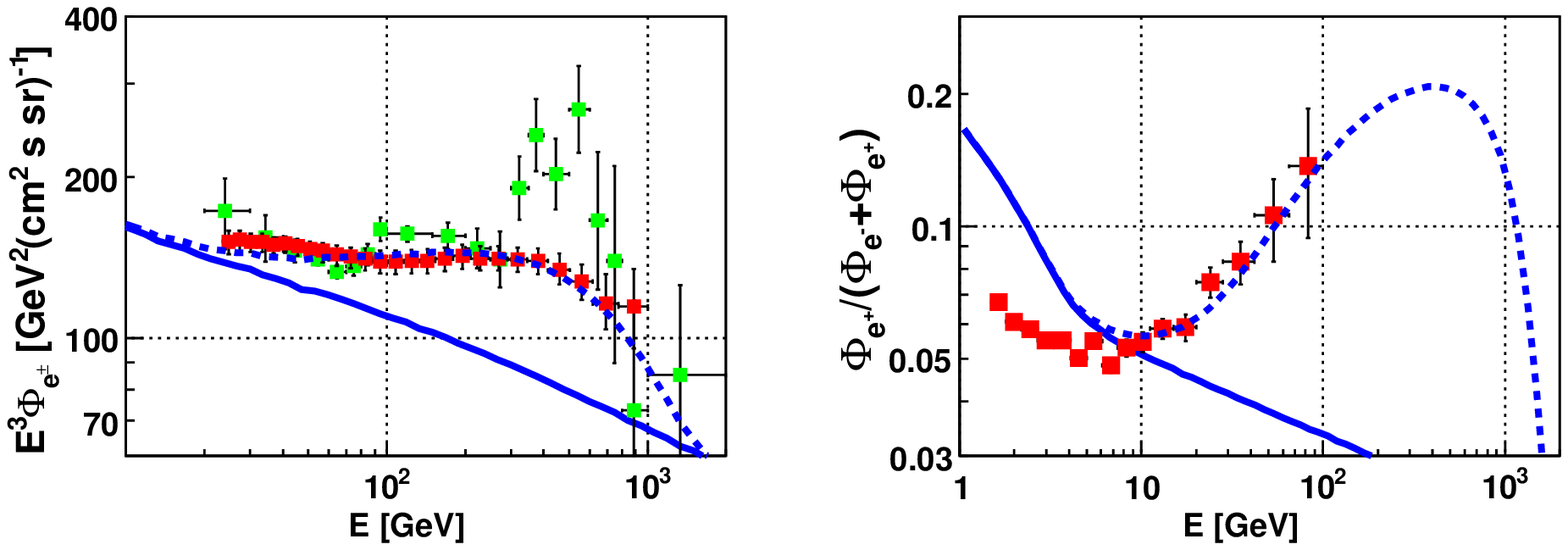}\\
\end{centering}
\caption{Left panels: fit of $L_1L_3\bar E_3$ (top) and $L_2L_3\bar
  E_3$ (bottom) operators (dashed blue) and {\sc GALPROP} background
  (solid blue) to electron-plus-positron spectrum from Fermi LAT (red,
  with error bars). Also shown is ATIC data (green, with error
  bars). Right panels: data on positron fraction from PAMELA (red,
  with error bars) shown with {\sc GALPROP} background (solid blue)
  and the result of the fit to the Fermi LAT data for the $L_1L_3\bar E_3$
  and $L_2L_3\bar E_3$ operators (dashed blue).
}\label{Fig:1800,LLE133}
\end{figure}

Recently, several anomalies in the data on cosmic ray electrons and
positrons have been reported. PAMELA has reported an anomalous rise in
the positron fraction above 10 GeV \cite{PAMELAep} and ATIC
\cite{ATIC} as well as Fermi LAT \cite{Abdo:2009zk} have reported an
excess in electrons plus positrons at around 100--800 GeV. It is
believed that these anomalies require a new source of high energy
electrons and positrons, and the most commonly discussed possibilities
are pulsars and dark matter \cite{Fermiint}. Our aim is to see how
well these anomalies can be fitted with the decay products from
gravitino dark matter, with an emphasis on the Fermi LAT and PAMELA
data.

In order to study the cosmic ray flux expected in experiments, we use
PYTHIA 6.4 \cite{PYTHIA} to calculate the spectra of the decay
products. We then let GALPROP \cite{GALPROP} propagate the particles
through the galaxy and calculate the expected background. The GALPROP
model we use is a conventional diffusion model where we have rescaled
the resulting primary electrons by a factor 0.75 to make room for a
simultaneous fit to PAMELA and Fermi LAT. For the dark matter halo
density, we assume a NFW profile \cite{NFW} with parameters $r_c =
20$~kpc and $\rho_0 = 0.33$~GeV~cm$^{-3}$.

Operators giving rise to jets, i.e. $LQ\bar D$ and $\bar U\bar D\bar
D$ operators, will produce large numbers of electrons and positrons
through charged pions. The resulting spectra, however, are too soft to
simultaneously fit PAMELA and Fermi LAT. Moreover, the non-observation
of any excess in the antiproton data reported by
PAMELA~\cite{PAMELApbar} essentially excludes all attempts with $LQ\bar
D$ and $\bar U\bar D\bar D$ operators.

We focus therefore on the nine $LL\bar E$ operators. When attempting
fits with these operators, it becomes clear that the three operators
with an SU(2) singlet field of the electron type ($\bar E_1$) give a
spectrum that is too hard; in order to get both a sufficiently soft
electron-plus-positron spectrum and a large enough positron
contribution at PAMELA energies, tau flavour in the SU(2) singlet
component, $\bar E_3$, works better. However, this is only true for
single coupling dominance; if combinations of operators are
considered, additional good fits can be obtained.

For gravitino masses around 2 TeV, $L_1L_2\bar E_3$ and $L_1L_3\bar
E_3$ give good fits to both PAMELA and Fermi LAT. If the gravitino
mass is increased towards 4 TeV, $L_2L_3\bar E_3$ also gives a good
fit. The absence of electron flavour requires a large gravitino mass
in order to reach the high end of the Fermi LAT
data. Figure~\ref{Fig:1800,LLE133} shows two such fits; $L_1L_3\bar
E_3$ ($L_2L_3\bar E_3$) with gravitino masses of 1.8 (3.7) TeV, and
other sparticle masses set to 2 (6) TeV.
%\goodbreak
\newpage
\section{Gamma Rays}

\begin{wrapfigure}{r}{6.5cm}
\vspace{-1cm}
\begin{centering}
  \includegraphics[width=6.5cm]{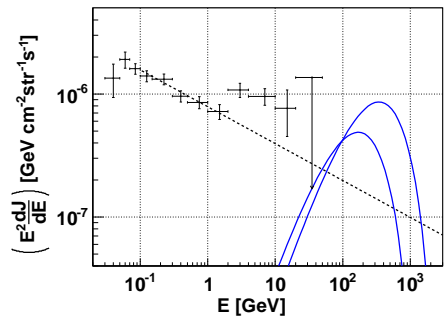}\\
\end{centering}
\caption{Photon spectra for the fits of the operators $L_1L_3\bar E_3$
  and $L_2L_3\bar E_3$ to the Fermi LAT data.  For comparison we also
  give the EGRET data on extragalactic diffuse emission
  \cite{reEGRET}.  }\label{Fig:gamma,LLE-133,233}
\vspace{-1cm}
\end{wrapfigure}

With a gravitino decay through an $LL\bar E$ operator, gamma rays
appear in two ways: internal bremsstrahlung off the produced leptons
and from the decay of mesons (mostly $\pi^0$) from $\tau$
decay. Figure~\ref{Fig:gamma,LLE-133,233} shows the resulting gamma
ray flux expected for experiments in the solar system, together with an extrapolation of the low-energy EGRET data. As one can see,
the gravitino masses that we consider are too large to be in conflict
with the EGRET data on extragalactic diffuse emission, but Fermi LAT
will eventually be able to either find some excess, or exclude this
model, for the parameter space under consideration.

\section{Conclusions}
Within R-parity violating SUSY models, gravitinos can be natural dark
matter candidates.  For trilinear R-parity violating operators of the
$LL\bar E$ type, the recent anomalies in cosmic ray electrons and
positrons can be accurately explained without contradicting other
cosmic ray measurements. This scenario will be tested in the near
future by the diffuse gamma ray data from Fermi LAT.

\end{document}